\pdfoutput=1
\documentclass[%
reprint,
 amsmath,amssymb,
 aps,
]{revtex4-1}
\usepackage{natbib}
\usepackage{graphicx}
\usepackage{dcolumn}
\usepackage{bm}
\usepackage{graphicx}
\usepackage{xcolor}

\begin{document}


\title{Experimental Observation of Nonreciprocal Waves in a Resonant Metamaterial Beam}

\author{M. A. Attarzadeh}
\author{J. Callanan}%
\author{M. Nouh}%
 \email{Corresponding author: mnouh@buffalo.edu}
\affiliation{Dept. of Mechanical \& Aerospace Engineering, University at Buffalo (SUNY), Buffalo, New York 14260-4400}%





\begin{abstract}
Space-time-varying materials pledge to deliver nonreciprocal dispersion in linear systems by inducing an artificial momentum bias. Although such a paradigm eliminates the need for actual motion of the medium, experimental realization of space-time structures with dynamically changing material properties has been elusive. In this letter, we present an elastic metamaterial that exploits stiffness variations in an array of geometrically phase-shifted resonators -- rather than external material stimulation -- to induce a temporal modulation. We experimentally demonstrate that the resulting bias breaks time-reversal symmetry in the resonant metamaterial, and achieves a nonreciprocal tilt of dispersion modes within dynamic modulation regimes.
\end{abstract}

\maketitle
%
The ability to control wave propagation in elastic media is of key importance in a number of disciplines that span multiple geometric scales. Resonant and periodic materials aim to provide a means to mitigate or guide elastic waves via precisely engineered periodic variations in structural geometry, allowing wave control features to scale with the structure itself \cite{doyle1989wave,mead1996wave,Hussein2014}. The field of elastic metamaterials possessing unique dispersive features including tunable band gaps, topological edge states, and negative effective properties has received increased attention in recent years \cite{Gonella_directivity,chen2014piezo,swinteck2015bulk,huang2009negative}. Most recently, novel configurations have been presented as pathways to break one of the fundamental elastodynamic principles, \textit{reciprocity}, and onset a diode-like behavior \cite{zanjani2014one,fleury2014sound,bilal2017bistable}. Reciprocity, often used in conjunction with principles of superposition and symmetry, is of key importance for many analysis methods in electromagnetism, acoustics, and signal processing -- in practice, however, back scattered waves present a number of issues and limitations in sensing, structural fidelity, telecommunication and defense applications. Therefore, it is of key interest to develop structures that exhibit a robust nonreciprocal wave propagation behavior. Linear systems with space-time modulated material fields have been the focus of a number of efforts to investigate wave amplification and nonreciprocity \cite{cullen1958travelling, cassedy1963dispersion}. Due to the time dependence of their properties, these systems are no longer bound by the reciprocity \cite{achenbach2012wave}, while their response remains amplitude-independent, unlike nonlinear counterparts. The theoretical problem of characterizing elastic wave propagation in space-time modulated systems has been investigated in one- \cite{trainiti2016non} and two-dimensional \cite{attarzadeh2018non} structures using plane wave expansion. Recently, this approach has been further extended to discretely modulated structures to better adapt with practical situations \cite{riva2019generalized}. Nonreciprocal waves have also been witnessed in locally resonant systems \cite{attarzadeh2018wave,nassar2017_meta} and elastic metamaterial beams \cite{nassar2017non}. Nonetheless, the realization of time modulated elastic systems remains a practical challenge due to their dynamic nature. A select number of efforts have conceived ways to realize nonreciprocal media. These include photo-elasticity \cite{swinteck2015bulk}, magneto-elastic effects \cite{Ansari17,wang2018observation,chen2019nonreciprocal}, piezoelectric materials \cite{croenne2017brillouin,trainiti2019time,marconi2019experimental}, geometric non-linearity \cite{wallen2019nonreciprocal,goldsberry2019non}, and gyroscopic action via angular momentum modulations \cite{JASAGyric}. Band gap tunability observed in granular materials with dielectric properties also provide promising avenues to break reciprocity \cite{nejadsadeghi2019frequency,nejadsadeghi2019axially}.
\begin{figure*}[]
     \centering
     \includegraphics[width=0.98\textwidth]{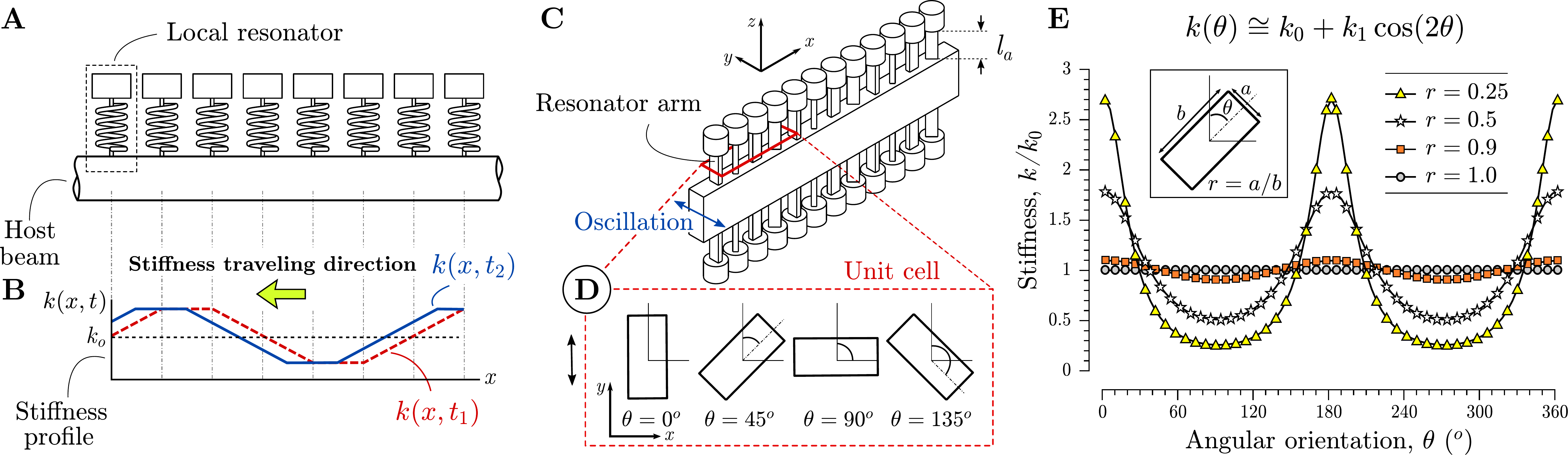}
     \caption{Schematic representation of the operational principles of the nonreciprocal metabeam: A. Illustration of a conventional metabeam with discretely located resonators; B. Space-time variation of resonators' stiffness traveling in the negative direction of the $x$-axis to induce artificial linear momentum bias; C. Proposed realization of the spatially modulated resonator stiffness in an elastic metabeam; D. Angular orientation of resonators in a unit-cell to create the spatial modulation; E. The stiffness variation of rectangular cross sections with various aspect ratios versus angular orientation.}
     \label{fig:Schematics}
\end{figure*}

\begin{figure}[h!]
    \centering
    \includegraphics[width=0.43\textwidth]{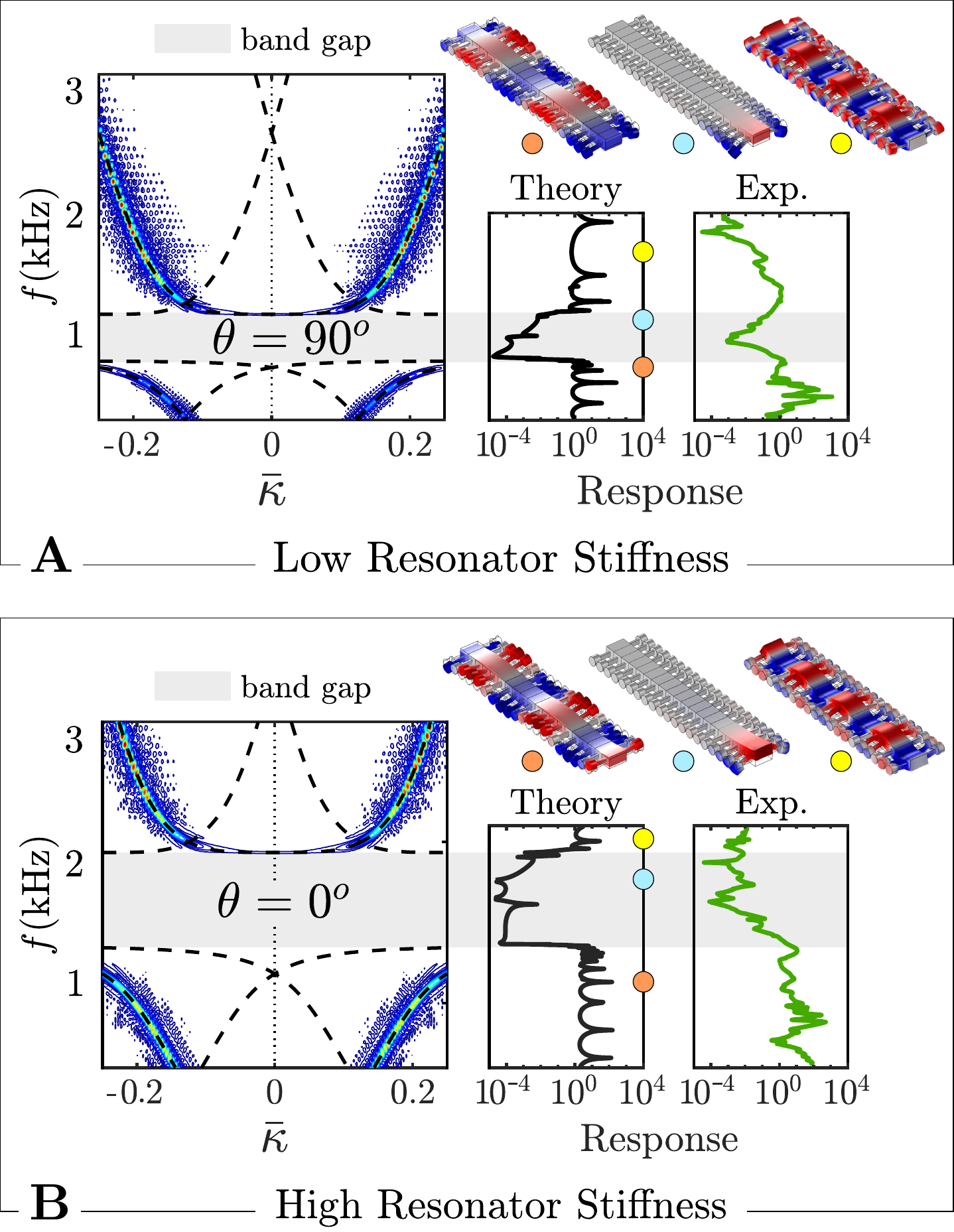}
    \caption{Dispersion behavior and transmission spectra of the metabeam in two \textit{non-rotating} configurations: A. Low stiffness (all resonators at $\theta=90^o$) and B. High stiffness ($\theta=0^o$). Shaded regions indicate band gaps. Mode shapes pre, within, and post both band gaps are provided for reference. Metabeam parameters identical to those listed in Fig.~\ref{fig:block}.}
    \label{fig:Quasi}
\end{figure}

In this work, we present a novel apparatus that achieves space-time modulation of elastic stiffness in a sub-wavelength configuration that does not involve smart or adaptive materials with inter-physical couplings. The system, which is designed and constructed using widely available components and manufacturing techniques, comprises a sub-wavelength elastic metamaterial beam (or metabeam, for short) and relies on local resonators that dynamically vary their effective stiffness by changing their angular orientation with respect to the vibration direction. Such a behavior has been recently used to tune locally resonant band gaps \cite{lv2019tunable}. In addition to breaking wave propagation symmetry, the proposed design inherits the tunability of frequency band gaps in conventional metamaterials, which in turn extends the nonreciprocal behavior to low frequencies. 
The nonreciprocal metabeam is conceptually akin to a locally resonant metamaterial as depicted in Fig.~\ref{fig:Schematics}a and its well-established dynamics \cite{sun2010,al2017mechanics}. The spring stiffness in each resonator, however, is varied independently in time, and by controlling the macroscopic spatial distribution of the resonators' stiffness, a space-time traveling profile is achieved. Fig.~\ref{fig:Schematics}b shows an example of the necessary effective stiffness variation to create nonreciprocity; the stiffness $k(x,t)$ is graphed for two time instants, $t_1$ and $t_2$. The curve $k(x,t_2)$ is identical to $k(x,t_1)$ except for a spatial phase shift. The means of achieving this stiffness variation are of paramount importance for both research and practical implementation purposes. 
Considering a beam with rectangular cross section (hereafter referred to as resonator arm) as the spring member, as shown in Fig.~\ref{fig:Schematics}c, the lateral stiffness under Euler-Bernoulli theory is given by $k = {3 E I_x}/{l^3}$, where $E$ is the elastic modulus, $l$ is the arm length, and $I_x$ is the second area moment of arm cross section calculated perpendicular to the vibration direction. We propose that the stiffness variation can be achieved by making use of the second area moment of the resonator arm instead of controlling its elastic modulus $E$. For a resonator rotated by the angle $\theta$, the value of $I_x$ can be computed by using the principle axes values and a coordinate rotation: let $x'$ and $y'$ be the principle axes for one resonator arm; the relevant moment of area for vibration in the $y$ direction as depicted in Fig.~\ref{fig:Schematics}c is $I_x = I_0 + I_1 \cos (2 \theta)$, where $I_0 = \frac{1}{2}(I_{x'}+I_{y'})$ and $I_1 = \frac{1}{2} (I_{x'}-I_{y'})$. Note that the difference in magnitude between the two principle moments is responsible for the alternating variation of $I_x$ and in turn the variation in the resonator stiffness. A series of finite element numerical simulations were carried out to verify the change in stiffness as a result of the resonator angular rotation. Fig.~\ref{fig:Schematics}e shows the stiffness variation with respect to the angular orientation for rectangular cross sections with aspect ratios $r = 0.25, \ 0.5, \ 0.9$ and $1$. Fig.~\ref{fig:Schematics}e confirms that a lower aspect ratio results in a greater variation in the arm stiffness as it rotates; the stiffness variation starts deviating from a perfect harmonic variation due to violation of Euler-Bernoulli beam assumptions at lower aspect ratios. The dependence of band gaps on resonator orientation is shown in literature \cite{lv2019tunable} and confirmed by studying the metamaterial in two different non-rotating configurations, as depicted in Fig.~\ref{fig:Quasi}. Dispersion diagrams and the corresponding transmission spectra confirm that energy propagation in the metabeam can be dramatically altered by changing only the orientation of the resonators by $90^\circ$ to switch from the lowest to the highest stiffness configurations (a band gap shift of nearly $400$ Hz). Matching experimental results, obtained from the setup outlined in the subsequent section, are also provided.
\begin{figure*}[t]
    \centering
    \includegraphics[width=0.75\textwidth]{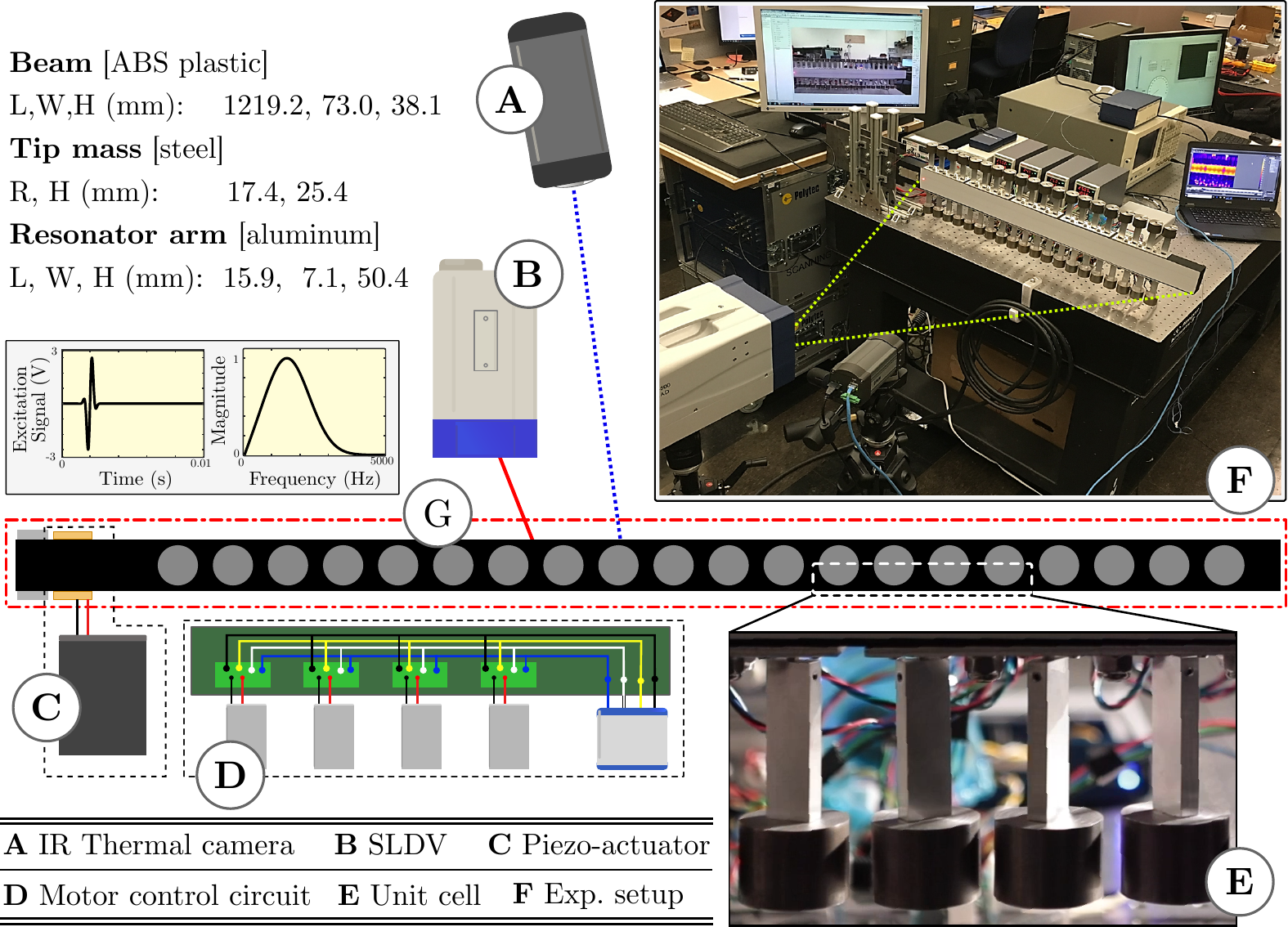}
    \caption{Illustration of the complete experimental apparatus for the nonreciprocal elastic metabeam: A. Thermal imaging camera; B. Scanning Laser Doppler Vibrometer (SLDV) imaging system; C. Beam excitation module and the signal sent to the piezoelectric actuator; D. Power and controller circuit; E. Unit cell (bottom half, close-up); F. and G. Metabeam.}
    \label{fig:block}
\end{figure*}

The host beam in Fig.~\ref{fig:Schematics}c is equipped with a series of local resonators that each consist of a prismatic arm and a tip mass. Fig.~\ref{fig:Schematics}d shows the angular orientation of the resonators in one unit cell in a metabeam with spatially modulated resonator stiffness. Each resonator is rotated by $45^\circ$ relative to the previous one such that a repeating unit cell is made up of four local resonators. Let the index $j$ denote a single resonator (more specifically a pair of resonators at the same $x$ location on the beam, top and bottom); a unit cell on the metabeam consists of $J$ resonator pairs. As such, a prescribed phase shift between the resonators angular orientation ($45^{\circ}$ adjacent resonators) generates a spatial modulation of the stiffness. 
The space-time modulation of the resonators' stiffness is achieved by a synchronized rotation of the resonators' arms with an angular velocity of $\omega_p$ while maintaining the aforementioned spatial modulation. The combined effect of both spatial and temporal variation induces the desired wave-like stiffness pumping. The stiffness of the $j^{\text{th}}$ resonator at time $t$ is $k^{(j)}(t) = k_0 + k_1 \cos (\omega_p t + \frac{2 \pi j}{J})$. 



%
\begin{figure*}[!tbp]
\vspace{1cm}
  \centering
  \begin{minipage}[b]{0.49\textwidth}
    \includegraphics[width=\textwidth]{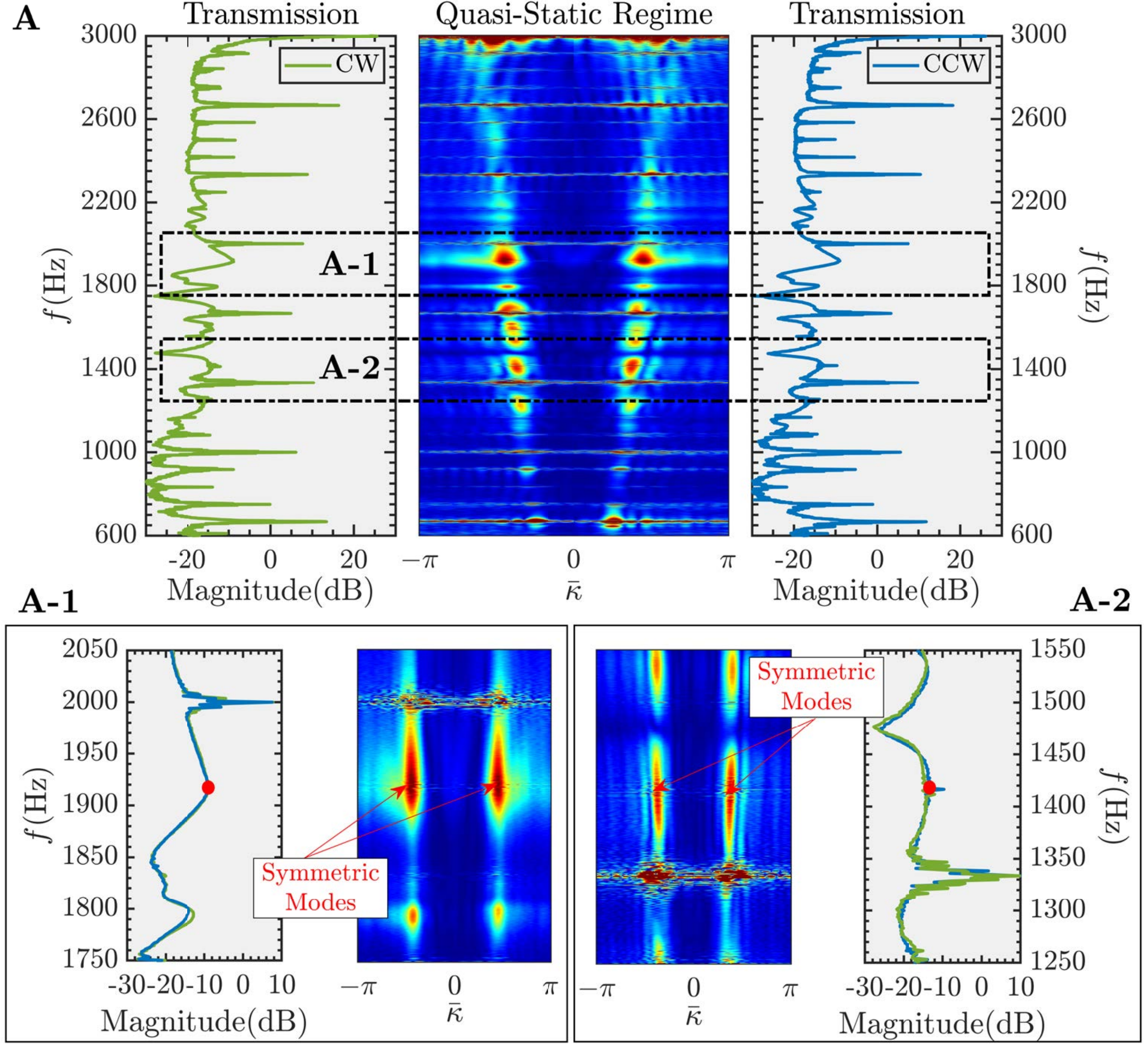}
  \end{minipage}
  \hfill
  \begin{minipage}[b]{0.49\textwidth}
    \includegraphics[width=\textwidth]{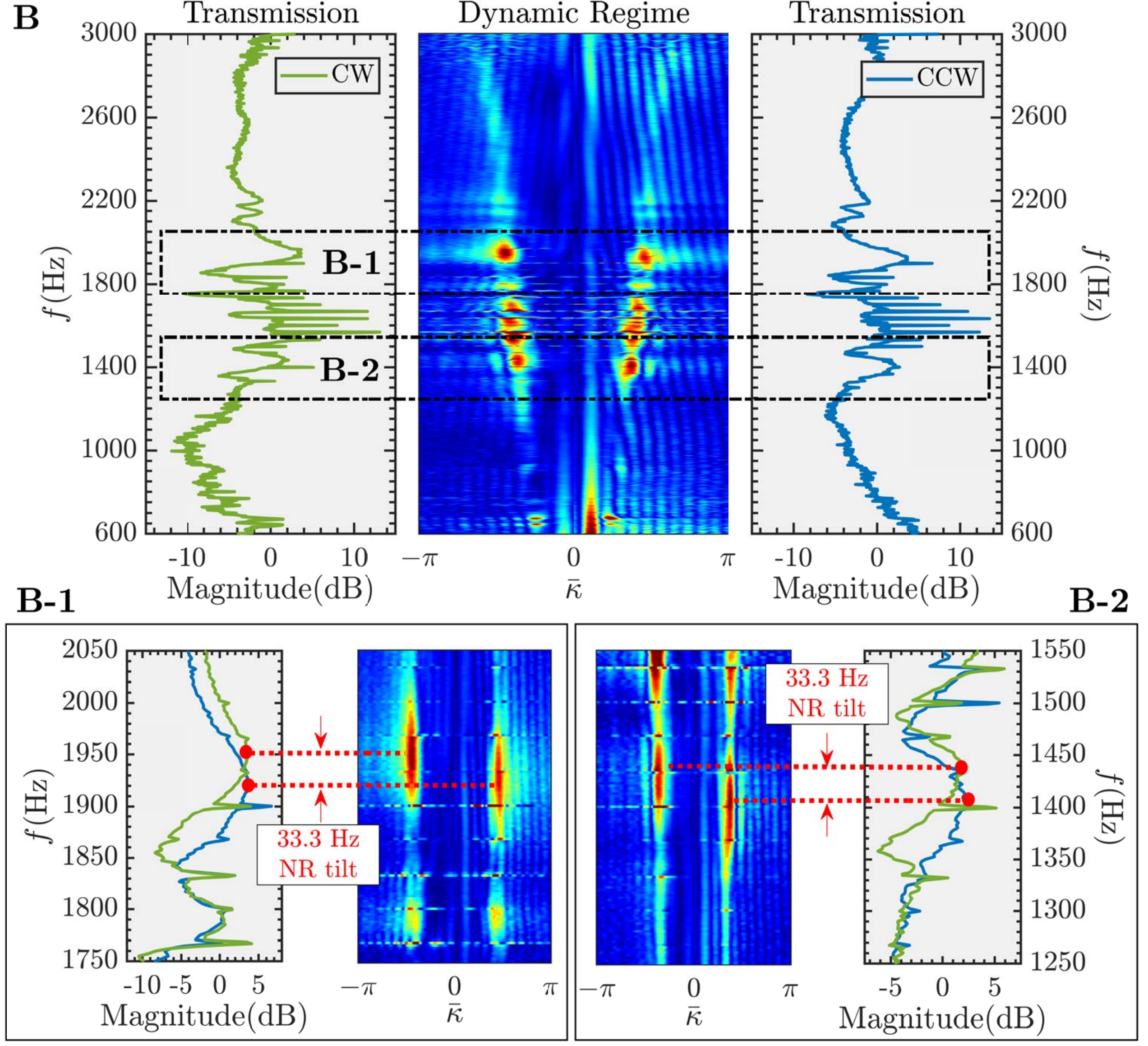}
  \end{minipage}
\caption{Experimental results:
A. Quasi-Static modulation regime with motors rotating at $100$ rpm (1.67 Hz). Experimentally reconstructed dispersion patterns and transmission spectra obtained from backward (CCW rotation) and forward (CW rotation) modulated structures showing reciprocal response with fairly symmetrical counter-propagating modes due to low speed modulation (A-1. Close-up: $1750-2050$ Hz. A-2. Close-up: $1250-1550$ Hz). B. Dynamic modulation regime with motors rotating at $2000$ rpm ($33.3$ Hz). Experimentally reconstructed dispersion patterns and transmission spectra obtained from backward (CCW) and forward (CW) modulated structures showing nonreciprocal tilt equal to the exact amount of the motors' rotational speed for counter-propagating modes (B-1. Close-up: $1750-2050$ Hz. B-2. Close-up: $1250-1550$ Hz).}
\label{fig:Regimes}
\end{figure*}

The metabeam was constructed following the general operating principles as shown in Fig.~\ref{fig:Schematics}. The setup consists of the host beam, forty local resonators grouped into symmetric pairs above and below the beam, and motors to control the resonator angle, along with the measurement and excitation systems, as illustrated in Fig.~\ref{fig:block}. The metabeam is represented by a black rectangle with the local resonators as circles, similar to a bird's eye view of the actual apparatus, in Fig.~\ref{fig:block}g. The stepper motors were controlled with a custom driver array (Fig.~\ref{fig:block}d). Vibration measurements were taken with a Polytec PSV-500 Scanning Laser Doppler Vibrometer (SLDV), illustrated in Fig.~\ref{fig:block}b. The temperature of the beam was continuously monitored with a FLIR A325sc thermal imaging camera (Fig.~\ref{fig:block}a) to ensure that the stiffness of the host structure did not vary due to heating or cooling. Actuators in the form of two extender piezoelectric plates (and a high voltage amplifier, Fig.~\ref{fig:block}c) were mounted symmetrically on both sides of the beam and operated $180^\circ$ out of phase such that transverse vibrations were dominantly excited. The motors and piezoelectric actuators were directly controlled using LabVIEW and an NI DAQ. Further, the SLDV system was triggered using the same controller in order to ensure that the measurements were synchronous with the prescribed temporal modulation of the metabeam. A close-up view of one unit cell in the metabeam is shown in Fig.~\ref{fig:block}e revealing the angular phase shifts between adjacent resonators in one cell. The time domain signal that was sent to the high voltage amplifier was a wide-band tone burst excitation with a central frequency of $1500$ Hz. 
For more details on the experimental set up, operation, and measurement synchronization, see the Supplemental Material and accompanying multimedia file.
Spatial and temporal Fourier transformations are performed on the SLDV velocity field data to extract the frequency content of propagating waves, and the result is then normalized by the excitation spectrum. By adopting a cantilever beam configuration with the ability to reverse the motor direction, we can effectively double the length of the beam without increasing the size of the structure: waves traveling from the piezoelectric actuator while the motors are rotating clockwise (CW) can be treated as waves traveling in the opposite direction while the motors are rotating counter-clockwise (CCW). 
The metabeam was tested at two different modulation regimes, \textit{quasi-static} and \textit{dynamic}, based on the rotational speed of the motors, for both CW and CCW rotations. The Supplemental Material multimedia file shows the apparatus in both operational regimes. 

In the quasi-static modulation regime, the motors were rotated at a relatively low speed, $100$ rpm ($1.67$ Hz), compared to the natural frequencies of the resonators -- which are higher than $1000$ Hz.
In addition, the resonators were oriented in a way to realize a spatially periodic variation of stiffness along the structure (similar to what is shown in Fig.~\ref{fig:Schematics}c). The combination of the spatial phase shift and the low-speed motor rotation generates space-time modulation of stiffness that slowly creeps along the metabeam. With a CCW direction command, a backward modulation appears that is moving from the end to the root of the beam, while a CW command results in a forward modulation. This slow variation of stiffness is reminiscent of adiabatic pumping in quantum mechanical systems \cite{kraus2012topological, wang2013topological}. Despite the proven unconventional topological aspects of such systems \cite{chaunsali2016stress,attarzadeh2018wave}, here we limit our attention to the spectral properties of the metabeam and dispersion characteristics rather than topological features of the modes. The metabeam was tested in both CW and CCW rotations at $100$ rpm and Fig.~\ref{fig:Regimes}a illustrates the experimentally retrieved dispersion and transmission of the metabeam within the frequency range of interest, $600$ to $3000$ Hz. The right half-plane of the dispersion contours corresponds to the CCW rotation and the left half-plane corresponds to the CW rotation, which are shown along with their respective transmissions in Fig.~\ref{fig:Regimes}a.
Note that the repeated flat lines in the dispersion plots (and alternating peaks in the spectra) are  artifacts of the stepping frequency of the stepper motors that occurs at $200$ times the rotational frequency (or $333.3$ Hz); Bipolar stepper motors with $1.8^\circ$ advance per step impart both the main harmonic and 200 times its frequency. These lines are kept in the results to maintain originality and reproducibility of the data. Due to their reciprocal nature, the motor vibration artifacts do not interfere with our interpretation of the results, as they perfectly match in CCW and CW dispersion patterns regardless of the modulation speed. From the zoomed view in Figs.~\ref{fig:Regimes}a-1 and ~\ref{fig:Regimes}a-2 we conclude that insignificant nonreciprocity between the forward and backward dispersion modes exists for this case. As intended, the traveling speed of the modulation (or the speed of motors) in the quasi-static modulation regime is insufficient to instigate detectable nonreciprocal response regardless of the rotational direction of the motors.

In the dynamic modulation regime, we intentionally increase the rotational speed of the motors to 2000 rpm (33.3 Hz) while maintaining the spatial modulation (angular position phase shift between the successive resonators). The result is a much faster traveling modulation and is shown in Fig.~\ref{fig:Regimes}b along with the transmission spectra for both forward and backward modulations. As observed from the zoomed view in Figs.~\ref{fig:Regimes}b-1 and ~\ref{fig:Regimes}b-2, the faster modulation speed in the dynamic regime generated one-way modal transition and detectable nonreciprocal tilt of the dispersion modes. As a result, the rightward propagating branch is down-shifted by the amount of 33.3 Hz (equal to the modulation speed, or rotational speed of motors) compared to the left propagating branch, which indicates the effectiveness of the proposed metabeam to break time-reversal symmetry.
In accordance with the previous findings on reciprocity breakage in space-time modulated systems the frequency shift between the forward and the backward propagating branches is an integer multiple of the modulation frequency \cite{chaunsali2016stress,nassar2018quantization}. Accordingly, the nonreciprocity can be further accentuated by increasing the rotational speed of the motors. The one-way transmission at a given frequency can also be switched in the opposite direction by simply rotating motors in the opposite direction.

This work introduced a first experimental realization of dynamically modulated metamaterials that do not rely on material response to external stimulus, but rather an inherent geometric attribute by design. A cornerstone feature of flexural materials, second area moment, was used to geometrically induce a space-time modulation and, consequently, enforce an artificial linear momentum bias to break time-reversal symmetry of waves in an elastic beam. It was experimentally demonstrated that nonreciprocal tilt of the dispersion modes is directly correlated with the modulation speed of the medium and can be adjusted all the way from complete reciprocal dispersion in the quasi-static regime to a complete nonreciprocal dispersion in the dynamic modulation regime. Nonreciprocal metamaterials can have extensive applications ranging from back scatter-free ultra-sonic imaging and sensor-actuator protection to duplex underwater communications, SONAR devices and nonreciprocal acoustic phased array radars. Further, the current design also shows great potential to harbor topologically interesting features in future efforts. For instance, the quasi-static regime allows for adiabatic evolution of eigen-structure which provides a realizable route to create topologically protected boundary modes in elastic structures.

The work was supported by the US National Science Foundation through Award no. 1847254 (CAREER), the Vibration institute, as well as the NY State Center of Material Informatics.

%


\end{document}